\documentclass[]{elsart}

\usepackage{graphicx}
\usepackage{amssymb}
\usepackage{epsfig}

\begin{document}

\begin{frontmatter}

\title{\bf An interest rates cluster analysis}
\author[label1,cor1]{T. Di Matteo}
\corauth[cor1]{Corresponding author:Tel: +61 2 61250166, fax: +61 2 61250732.}
\ead{tiziana.dimatteo@anu.edu.au},
\author[label1]{T. Aste},
\author[label2]{R. N. Mantegna}
\address[label1]{Applied Mathematics, Research School of Physical Sciences, Australian National University, 0200 Canberra, Australia.}
\address[label2]{INFM-Dipartimento di Fisica e Tecnologie Relative, Universit\`a di Palermo - viale delle Scienze, 90128 Palermo, Italy.}

\begin{abstract}
An empirical analysis of interest rates in money and
capital markets is performed. We investigate a set of $34$ different weekly
interest rate time series during a time period of 16 years 
between $1982$ and $1997$.
Our study is focused on the collective behavior of the stochastic fluctuations
of these time-series 
which is investigated 
by using a clustering linkage procedure. Without any a
priori assumption, we individuate a meaningful separation in 6 main clusters
organized in a hierarchical structure. 
\end{abstract}


\end{frontmatter}

\section{Introduction}

Since long time financial data have been widely studied by economists,
mathematicians and, more recently, by physicists \cite{LibrMant,BouchaudPot,LibDac}. The variations of these financial time series can be seen as stochastic processes where a set of financial quantities is varying in time as consequence of underlying economic changes. The present availability of enormous sets of financial data allows to get increasingly important insights on the complex behavior of these systems starting from 
empirical studies. These investigations are leading to more and more accurate results on risk assessment and search for market imperfections. One of the important points in these analyses is to individuate similarities and specificities among the analyzed financial time series. This search has been widely exploited for stocks price changes whereas interest rates have been less investigated \cite{Pagan,Rebonato,Bouchaud,Bernaschi2002,DiMatteo,Nuyts}. For several economic reasons, interest rates and bonds have very similar statistical behavior in time or, in other words, they are all highly correlated . Their multivariate dynamics has been studied with a correlation-based clustering procedure in a set of US treasury securities where an underlying 
hierarchical structure has been detected \cite{Bernaschi2002}. Here we investigated
a partially different and less homogeneous set to evaluate the 
degree of hierarchal organization among different time series
in a diversified group of bonds. 

\section{An empirical analysis on interest rates} 
\label{s.emp}
We investigate weekly data for 34 selected interest rate time series  recorded
in the Federal Reserve (FR) Statistical Release database \cite{data}. In the following we
will indicate these time series with the symbol $f_{i}(t) $, where $t$ is the
current date and $i$ is a number which labels the different time series (see
Tab.\ref{t.1}).
\begin{table}
\caption{Interest rates and standard deviations in the time period 1982-1997.}
\label{t.1}
\begin{tabular}{cccccccccccc}
$ i $ & $f_i$ & $\sigma_i$ &$ i $ & $f_i$ & $\sigma_i$ &$ i $ & $f_i$ & $\sigma_i$ &$ i $ & $f_i$ & $\sigma_i$ \\
\hline
1 & FED & 0.30935 & 10 & BA6 & 0.17225 & 19 & TC5 & 0.15715  & 28 & TC10P& 0.13608 \\
2 & SLB & 0.13001 & 11 & CD1 & 0.21291 & 20 & TC7Y & 0.15363  & 29 & ED1M & 0.2166 \\
3 & CP1 & 0.22257 & 12 & CD3 & 0.1901 & 21 & TC10Y & 0.14863 & 30 & ED3M & 0.19926 \\
4 & CP3 & 0.19011 & 13 & CD6  & 0.19299 & 22 & TC30Y & 0.13288 & 31 & ED6M & 0.20104\\
5 & CP6& 0.17951  & 14 & TC3M & 0.1925 & 23 & TBA3M & 0.21838 & 32 & AAA & 0.10695 \\
6 & FP1 & 0.21418 & 15 & TC6M & 0.18271 & 24 & TBA6M & 0.20186 & 33 & BAA & 0.09411\\
7 & FP3  & 0.15407 & 16 & TC1Y & 0.16993 & 25 & TBS3M & 0.17672 & 34 & CM & 0.11556\\
8 & FP6 & 0.13842 & 17 & TC2Y & 0.16347 & 26 & TBS6M & 0.16262 \\
9 & BA3 & 0.17838  & 18 & TC3Y & 0.16227 & 27 & TBS1Y & 0.14789 \\
\end{tabular}
\end{table}
The different interest rate time series analyzed are: The Federal funds rate (FED); State $\&$ local bonds (SLB); Commercial Paper (CP); Finance Paper placed directly (FP); Bankers acceptances (BA); The rate on certificates of deposit (CD); (Note that in these cases the numbers $1$, $3$ and $6$ stand for maturity dates of $1$, $3$ and $6$ months.); The yields on Treasury securities at `constant maturity' (TC) (in particular the TC at 3 and 6 months (TC3M, TC6M) and 1, 2, 3, 5, 7, 10, and 30 years (TC1Y-TC30Y) maturities; The Treasury bill rates (TBA) with maturities of 3 and 6 months (TBA3M, TBA6M, TBS3M, TBS6M) and 1 year (TBS1Y); The Treasury long-term bond yield (TC10P); The Eurodollar interbank interest rates (ED) with maturity dates 1 month, 3 months and 6 months (ED1M, ED3M, ED6M), respectively; The Corporate bonds Moody's seasoned rates (AAA, BAA) and The Conventional mortgages rates (CM). Their characteristics can be found in \cite{data}. Unless differently stated, we report weekly data obtained from unweighted averages of daily data ending on Friday.

\section{Fluctuations}
\label{s.prob}
\begin{figure}
\begin{center}
\begin{tabular}{cc}
\mbox{\epsfig{file=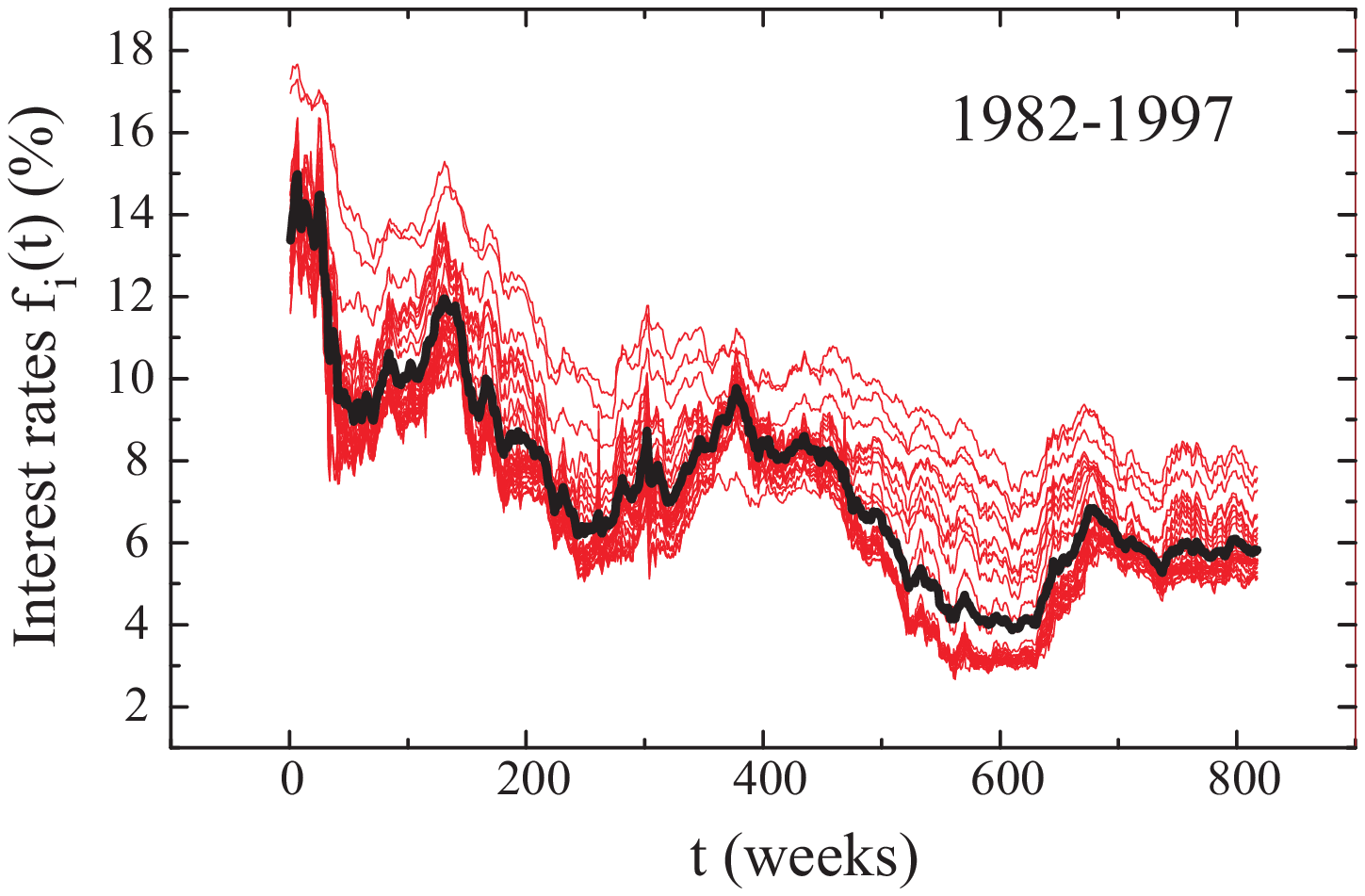,width=6.cm,angle=0}}
&\mbox{\epsfig{file=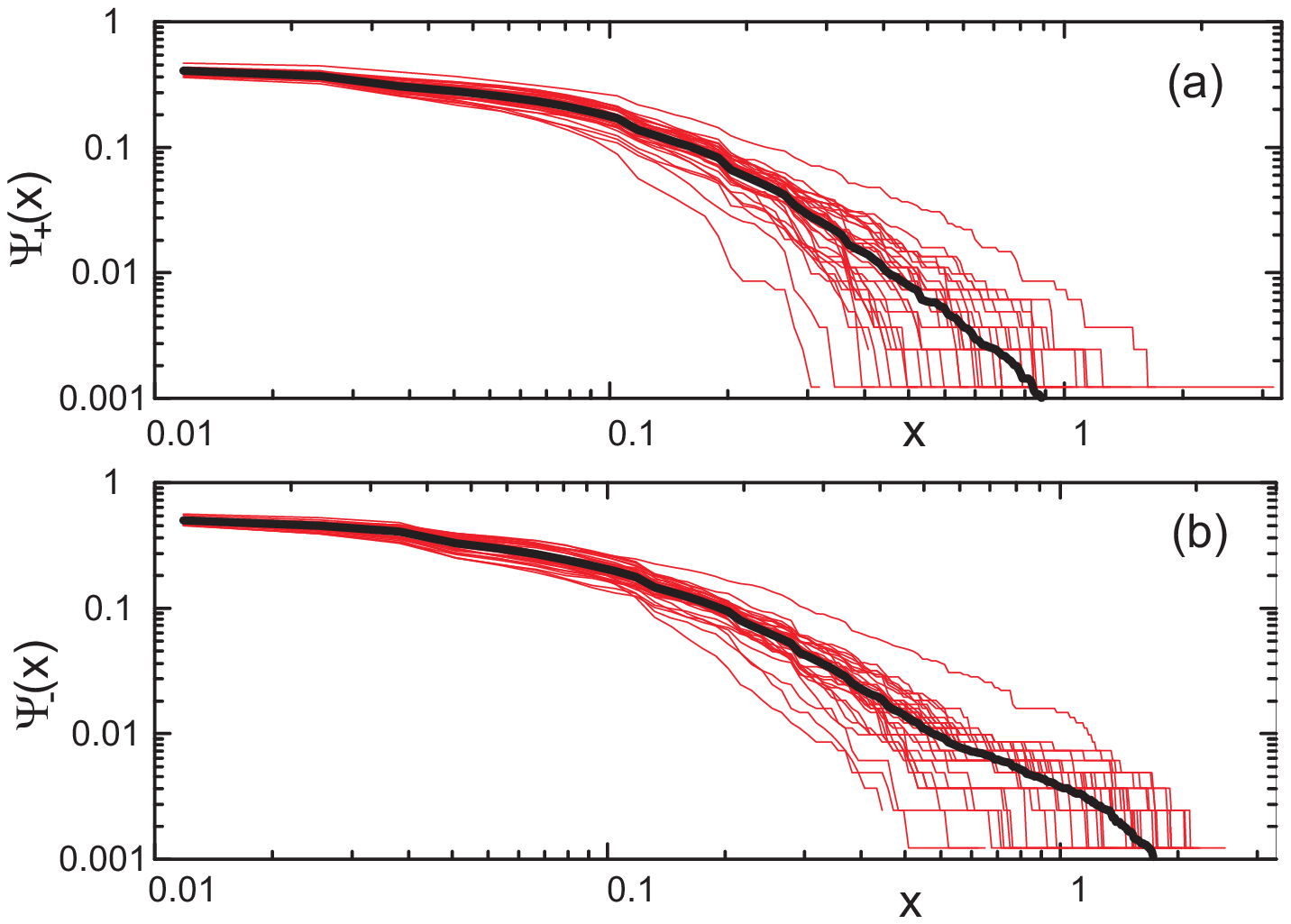,width=6.cm,angle=0}}
\end{tabular}
\end{center}
\caption{Left) Interest rates $f_i(t)$ as function of $t$ for $i=1-34$
(grey lines) and their average $\bar f(t)$ (black line). Right) Cumulative distributions for the probabilities of the interest rates fluctuations. (a) $\Psi_+(x)$ and (b) $\Psi_-(x)$. (The black lines are their averages.)}
\label{f.allFt}
\end{figure}

\begin{figure}
\begin{center}
\begin{tabular}{cc}
\mbox{\epsfig{file=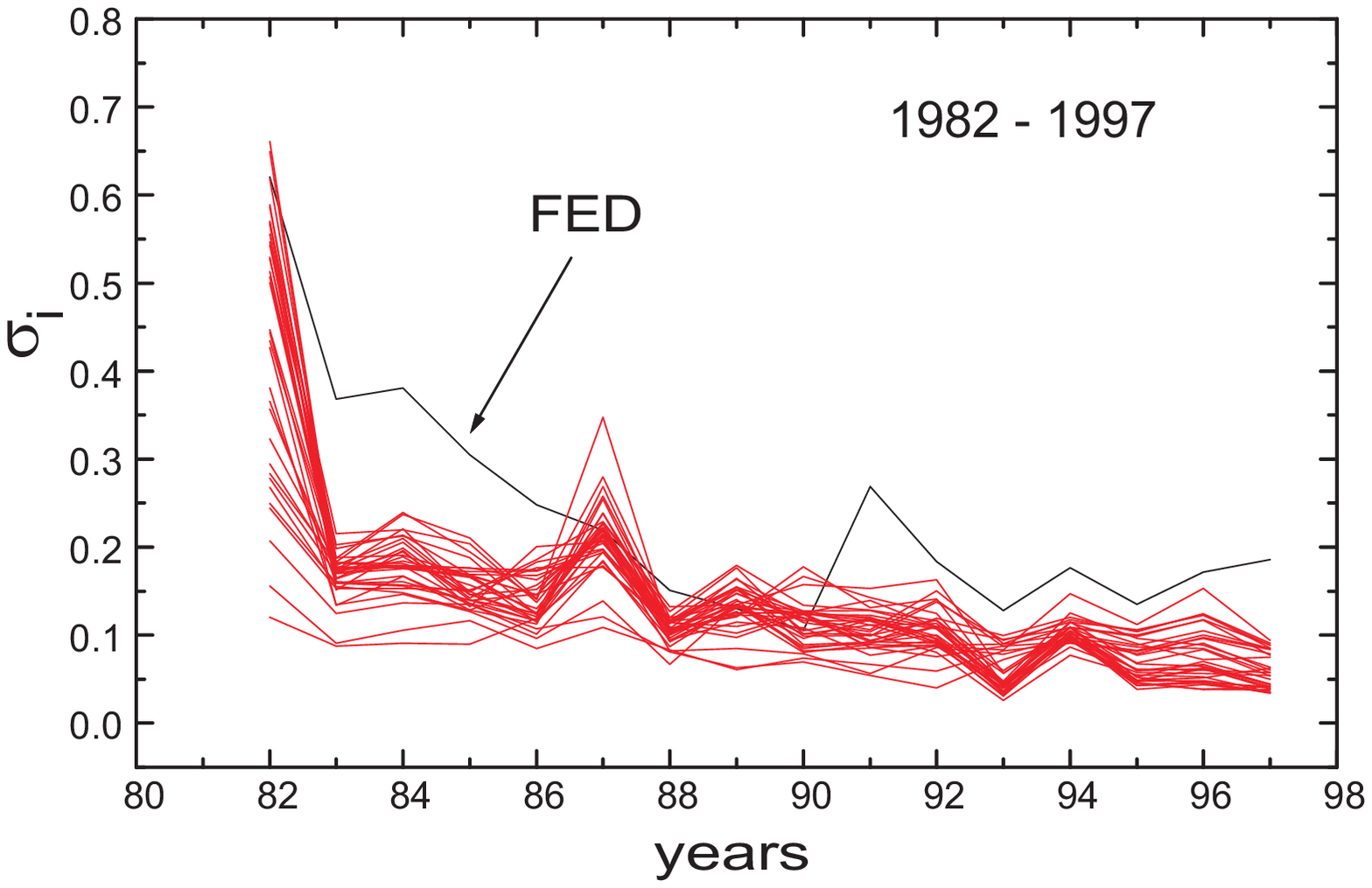,width=6.cm,angle=0}}
&\mbox{\epsfig{file=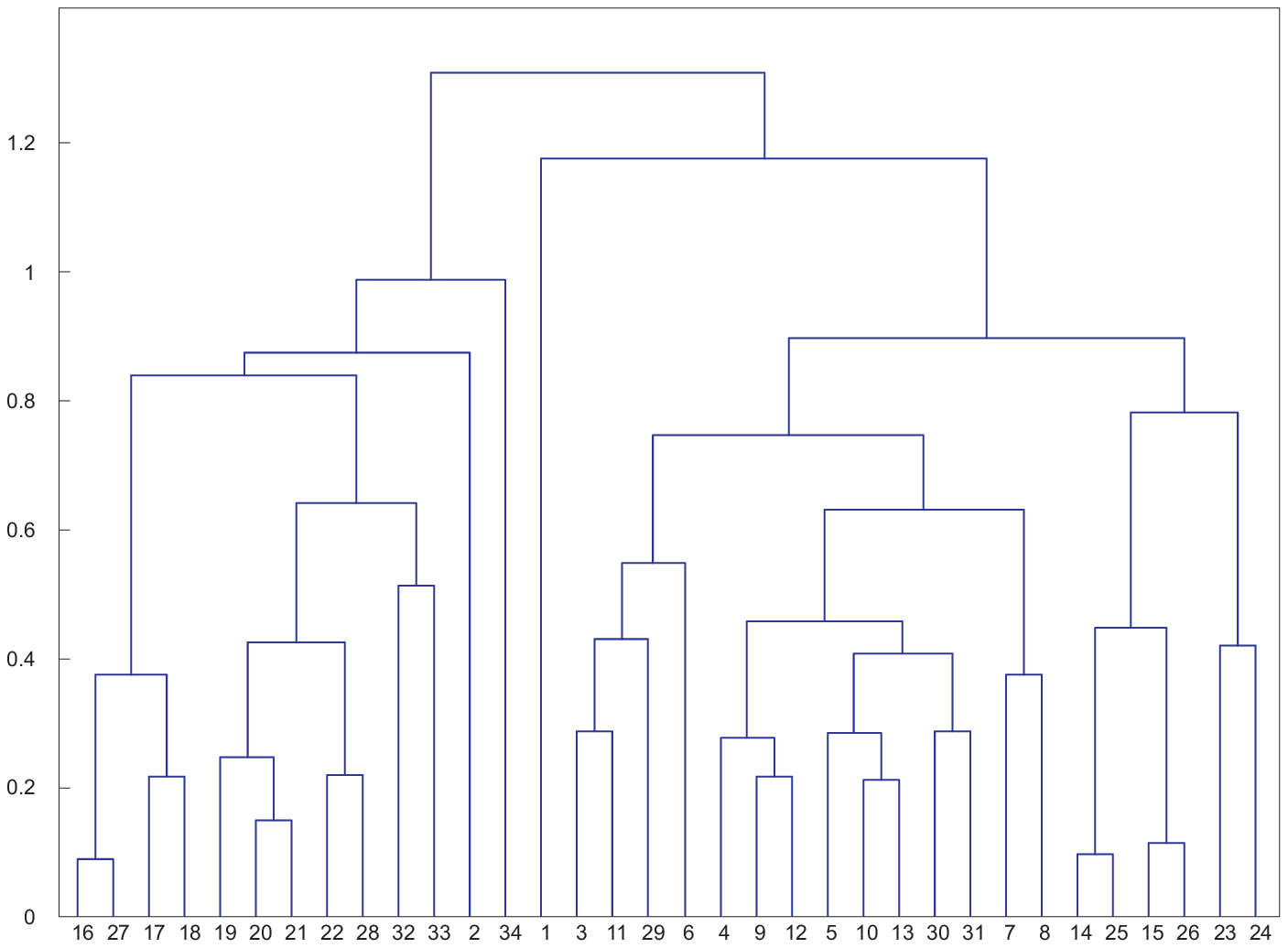,width=6.cm,angle=0}}
\end{tabular}
\end{center}
\caption{Left) Standard deviations of $\Delta f_i(t)$ for all the rates analyzed ($i=1..34$) as function of the years between 1982-1997. Right) Hierarchical tree obtained from the correlation coefficients of the 34 interest rates fluctuations time series $\Delta f_i (t)$ in the time period 1982-1997. (On the $x$-axis are reported the $i$ values and on the $y$-axis the ultra-metric distances.)}
\label{f.sigma}
\end{figure}

The interest rate time series, $f_{i}(t)$ v.s. $t$ are shown in Fig.\ref{f.allFt}, where their average $\bar f(t) = \sum_i f_i(t)/34$ is also
shown. It is evident from Fig.\ref{f.allFt} that all
these data follow very similar trends in time and they lay in a narrow band
around $\bar f(t)$.
The interest rate fluctuations are analyzed by studying the changes in their values from one week to the following week: $\Delta f_{i}(t) = f_{i}(t+\Delta t) - f_{i}(t)$
where $\Delta t=1$ week. The quantities $\Delta f_{i}(t)$ show stochastic fluctuations around the zero with similar behaviors for all the interest rates.
These fluctuations are analyzed in the `tails' region by computing the cumulative distributions $\Psi_\pm(\pm x)$ ($x>0$), a quantity which tells us the probability to find a weekly change which is \emph{larger than} $x$ ($+$),
or \emph{smaller than} $-x$ ($-$).
It is defined as: $\Psi_+(x) = 1 - {\int^{x}_{-\infty}} { p(\xi) d \xi}$ and $\Psi_-(x) = {\int^{-x}_{-\infty}} { p(\xi) d \xi}$
with $p(\xi)$ being the probability density distributions of $\Delta f_{i}(t)$ (see Fig.\ref{f.allFt} (Right)). These distributions are highly leptokurtic and are characterized by non-Gaussian profiles. The standard deviation of $\Delta f_{i}(t)$ is defined as: $\sigma_i = \sqrt {{1 \over T_2-T_1} {\sum_{t=T_1}^{T_2}}(\Delta {f_{i}(t)}- \left< \Delta f \right>)^2}$ where $T_1$ and $T_2$ delimit the range of $t$, and $\left<\Delta f \right>$ is the average over time of $\Delta {f_{i}(t)}$ (which tends to zero for $T_2-T_1 \rightarrow \infty$). We compute the standard deviations of $\Delta {f_{i}}$ for each interest rate series for the whole period 1982-1997 (see Tab.\ref{t.1}) and for each year (see Fig.\ref{f.sigma} (Left)). We can observe an overall decreasing trend of $\sigma_i$ in the time period 1982-1997 with similar fluctuations for all the interest rates series except for FED.

\section{Cluster analysis and discussion}
\label{s.cluster}
\begin{figure}
\begin{center}
\begin{tabular}{cc}
\mbox{\epsfig{file=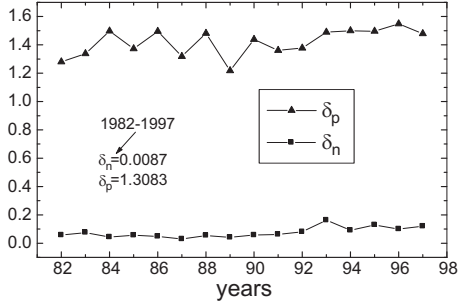,width=6.cm,angle=0}}
&\mbox{\epsfig{file=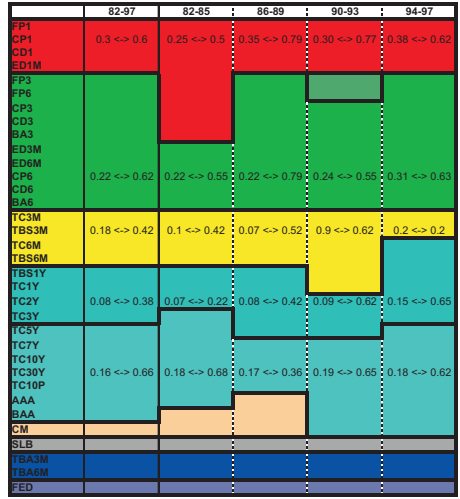,width=6.cm,angle=0}}
\end{tabular}
\end{center}
\caption{Left) Ultra-metric distances $\delta_n$ and $\delta_p$ at which the clustering process begins and ends, as function of the years between 1982-1997. Right) Cluster-structure persistence in the period 1982-1997. In the first column the interest rates are indicated. In the second column, the grey tones distinguish the different clusters as resulting from the analysis over the whole time period. The other columns refer to the cluster analysis over the four selected time periods, namely 82-85, 86-89, 90-93, 94-97. The numbers inside each cluster refer to the ultra-metric distances at which each cluster starts and ends its clusterization.}
\label{f.delta}
\end{figure}

To understand the geometrical and topological structure of the correlation
coefficients, we use the metric distance $d_{i,j}$ between
the series $\Delta f_i $ and $\Delta f_j$ which is defined in \cite{Gower66} 
and used for financial time series in \cite{RMant}: $d_{i,j} = {\sqrt{2(1-c_{i,j}) }}$ with
$c_{i,j}$ the correlations among the $i,j$ interest rates weekly changes:
\begin {equation}
c_{i,j} = {{ \left<\Delta f_i \Delta f_j\right> - \left<\Delta f_i\right> \left<\Delta f_j\right>} \over {\sigma_i \sigma_j} } \;\;\;,
\label{cij}
\end {equation}
where the symbol $\left<...\right>$ denotes a time average performed over the investigated time period.
The correlation coefficients are computed between all the pairs of indices labeling our interest series.
Therefore we have a $34 \times 34$ symmetric matrix with $c_{i,i}=1$ on the diagonal.
By definition, $c_{i,j}$  is equal to zero if the interest rates series $i$ and $j$ are totally uncorrelated, whereas $c_{i,j} = \pm 1$ in the case of perfect correlation/anti-correlation.
Therefore $d_{i,j}$ can vary between 0 to 2. We determine an ultra-metric distance $\hat d_{i,j}$ which satisfies the first two properties of the metric distance and replaces the triangular inequality with the stronger condition: $\hat d_{i,j} \leq max [ \hat d_{i,k},\hat d_{k,j}]$, called `ultra-metric inequality'.
Once the metric distance $d_{i,j}$ is used, one can introduce several ultra-metric distances. Mantegna et. al have used the `subdominant ultra-metric', obtained by calculating the minimum spanning tree connecting 
several financial time series \cite{Bonanno,Bonanno2001,Micciche2003,Bonanno2003}.
Here we consider a different ultra-metric space that emphasizes the cluster-structure of the data. In our case, a ``cluster'' is a set of elements with relative distances $d_{i,j}$ which are smaller than a given threshold distance $\bar \delta$, whereas disjoined clusters have some elements which are at distances larger than $\bar \delta$. We define the \emph{ultra-metric distance} $\hat d_{i,j}$ between two distinct elements $i,j$ belonging to two different clusters as the maximum metric distance between all the couples of elements in the two clusters \cite{DiMatteo}. The linkage procedure yields to a hierarchical graph as shown in Fig.\ref{f.sigma} (Right), which refers to the 34 time series in the whole period 1982-1997. The clustering process starts with a nucleation between TC1Y and TBS1Y at the ultra-metric distance $\delta_n=0.087$.
The clustering ends when all the series merge in a unique large cluster at the ultra-metric distance $\delta_p = 1.3083$. At this distance all the interest rates with maturity dates smaller or equal than 6 months (already merged with the FED at $\hat d=1.18$) make a single cluster with another cluster composed of interest rates with maturity dates larger or equal than 1 year.
The same analysis performed on each year, gives comparable $\delta_n$ and $\delta_p$ values which are plotted in Fig.\ref{f.delta} (Left).
Let now consider the intermediate region by analyzing the cluster evolution at the threshold distance $\bar \delta = 1/\sqrt{2}=0.707...$, which is half the way between completely uncorrelated series ($c_{i,j}=0$ and $d_{i,j} = \sqrt{2}$) and completely correlated ones ($c_{i,j}=1$ and $d_{i,j} = 0$).
At this threshold distance, the cluster analysis on the whole data set (1982-1997) leads to 6 clusters and 3 isolated elements, as one can see from Fig.\ref{f.sigma} (Right).

\begin{figure}
\begin{center}
\mbox{\epsfig{file=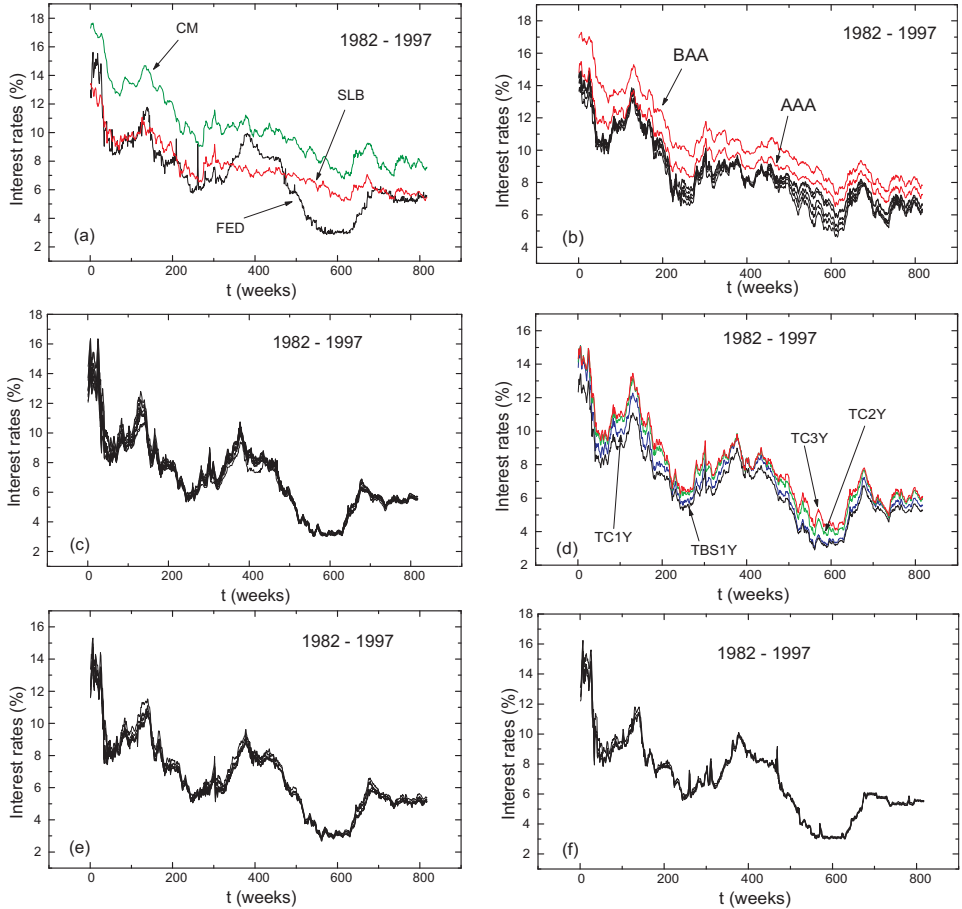,width=5.in,angle=0}}
\end{center}
\caption{Interest rates behaviors in the period 1982-1997. The figures refer to the data sets gathered into the clusters obtained from the linkage procedure. (a) FED, SLB and CM; (b) BAA, AAA, TC5Y, TC7Y, TC10Y, TC30Y, TC10P; (c) CP3, CP6, FP3, FP6, BA3, BA6, CD3, CD6, ED3M, ED6M;  (d) TC1Y, TC2Y, TC3Y, TBS1Y; (e) TC3M, TC6M, TBA3M, TBA6M, TBS3M, TBS6M; (f) CP1, FP1, CD1, ED1M.}
\label{f.FiClust}
\end{figure}

The corresponding interest rates associated with these clusters are summarized in first column (1982-1997) of Fig.\ref{f.delta} (Right) where the ultra-metric distances at which the nucleation process starts and ends for each cluster are also indicated. As one can see, the empirical analysis allow us to distinguish several different clusters that gather together meaningful quantities:
\begin{itemize}
\item[-] all the interest rates with maturities equal to 1 months;
\item[-] all the interest rates with maturities 3 and 6 months with distinctions for the Treasury securities at `constant maturity' (TC), Treasury bill rates (TBA) and Treasury bill secondary market rates (TBS);
\item[-] all the interest rates with maturities between 1 and 3 years;
\item[-] all the interest rates with maturities larger than 3 years.
\end{itemize}
Fig.\ref{f.FiClust} reports the plot for the interest rates time series $f_i(t)$ grouped into the different sets retrieved from the cluster analysis described above.
For some of the series the data-collapse is impressive, indicating that the correlations inside the clusters are strong in any part of the analyzed period.
It is therefore interesting to investigate whether a cluster structure, similar to the one obtained for the period 1982-1997, could be retrieved from an analysis on shorter periods.
This is of course a delicate point since the fragmentation of the data sets will increase the fluctuations due to the noise.
We choose to divide the whole period 1982-1997 in four smaller periods of 4 years. The results are reported in Fig.\ref{f.delta} (Right) where it is evident how the cluster structure is mostly conserved (and partially modified) in this 4-years period analysis. In conclusion, from the analysis of different kinds of interest rates in money and capital markets, referring to government, private, industries securities and commitments, we have shown how the used clustering linkage procedure is useful to detect differences and analogies among these tangled correlated data.

\end{document}